# Crystal Structure of $Rb_4C_{60}$ and $K_4C_{60}$ under Pressure

Ashfia Huq[1,2] and Peter W. Stephens[2]

1. Intense Pulsed Neutron Source, Argonne National Laboratory, Argonne, IL 60439 and

2. Department of Physics and Astronomy, State University of New York, Stony Brook, NY 11794

We show that $Rb_4C_{60}$ transforms from its ambient pressure orientationally disordered tetragonal structure to an ordered orthorhombic structure (isostructural to $Cs_4C_{60}$) at or above 0.4 GPa.  This opens the possibility of studying the previously observed metal-insulator transition under pressure in this narrow band system as a function of a continuous variable, rather than formulating distinct models for conducting and insulating fullerides.  No such transition is observed in $K_4C_{60}$ to 2.0 GPa

71.20.Tx, 71.30.+h, 61.10.-i, 61.50.Ks

## Introduction:

Alkali fullerides show a wealth of phenomena related to strong electron correlations and narrow bands, manifested by their close connections between crystal structure and electronic properties[1]. In the simplest view of $A_xC_{60}$ (where $1 \leq x \leq 6$ and A=Na, K, Rb and Cs), the electrons donated by alkali atoms are transferred to the six available $t_{1u}$ $C_{60}$ molecular orbitals rendering the material to be a conductor for $x < 6$. The conductivity of fcc $A_3C_{60}$ and $A_1C_{60}$ follows from this partially filled band. This picture is complicated by the fact that the electron-electron interaction $U$ on a single fullerene, on the order of 1 eV[2] is substantially larger than the band width $t$ (~ 0.5 eV). Gunnarson *et al.* have explained that the Hubbard $U$ should be renormalized down by the 3-fold orbital degeneracy so that the Mott Hubbard transition takes place for $U/t \sim \sqrt{3}$. As a result, the material is not a Mott insulator[3]. However, such a one-electron model fails to explain the mechanism leading to the insulating nature of $A_4C_{60}$ (A=K, Rb and Cs), which are insulating and non magnetic. (The exception, $Na_4C_{60}$, contains covalent inter-fullerene bonds[4]). $K_4C_{60}$ and $Rb_4C_{60}$ are both tetragonal, and so the most prosaic explanation for their insulating nature is that the crystal field splits the conduction band enough to open a gap. Detailed calculations do not support that hypothesis, even if they allow for Jahn-Teller distortion of the $C_{60}^{4-}$ ion, which would presumably enhance the stability of an insulating state[5]. There are several suggestions to explain how an insulating state could be produced by an even number of electrons, generally associated with a JT distortion[6]. However, the information experimentally available is not sufficient to distinguish among these, or to motivate a complete description of the insulating and conducting states of alkali fullerides in general.

Observations by Kerkoud *et al.* that the $^{13}C$ NMR relaxation rate in $Rb_4C_{60}$ changes from activated to Korringa (characteristic of the presence of conduction electrons) at a pressure of 0.8 GPa opened the possibility of studying a metal-insulator transition in that material[7], and thereby directly addressing the difference between conducting and insulating states of $A_4C_{60}$. In order to interpret that observation, it is first necessary to determine the crystal structure as a function of pressure. Accordingly, we have measured the crystal structure of $K_4C_{60}$ and $Rb_4C_{60}$ at pressures up to 2.0 GPa[8].

At ambient pressure, $K_4C_{60}$ and $Rb_4C_{60}$ phases both adopt tetragonal structures (Space group *I4/mmm*), with merohedral disorder similar to fullerenes in the superconductors. The high temperature form of $Cs_4C_{60}$ is isostructural with $K_4C_{60}$ and $Rb_4C_{60}$ but undergoes a structural change to an ordered orthorhombic (Space group *Immm*) phase below 623 K[9]. Evidence of a lower symmetry phase has also been seen in $K_4C_{60}$ at 250 K from infrared spectra[10]. The ground state of $Na_4C_{60}$ on the other hand forms a two-dimensional polymer and exhibits metallic conductivity[4]. Recent study of $Rb_4C_{60}$ structure under pressure has treated the structure as tetragonal from ambient pressure to 5.2GPa[11]. In our studies of $Rb_4C_{60}$ under pressure we find convincing evidence of a disorder-order transition at or above 0.4GPa.

**Experiment:**

Samples of $Rb_4C_{60}$ and $K_4C_{60}$ were prepared by solid-state reaction of commercially available high purity Rb or K metal and sublimed high purity $C_{60}$ powder. The sealed Pyrex tubes were heated between 200ºC and 500ºC for several months with intermediate grinding and adjustment of ratio in response to x-ray measurements. Minority phases of $Rb_3C_{60}$ and $K_3C_{60}$ were present in the final samples. The samples were loaded in an inert-atmosphere glovebox into the 500μm diameter hole of a stainless steel gasket of thickness 500μm between the two 600μm culets of a diamond anvil cell. The sample was mixed with Fluorinert as the pressure transmitting medium and a small grain of ruby to monitor the pressure based on the shift of optical fluorescence lines. After preliminary data at National Synchrotron Light Source beamline X3B1, we collected data on a different sample at undulator beamline ID-6 at the Advanced Photon Source. The cell was rocked by 0.5º to improve powder statistics. We limited the data analysis for the fullerene samples to *d* spacing of 2.375 Å since the diffraction peaks from the gasket, ruby, and Be window of the cell overwhelmingly dominate the pattern beyond this range. Refinements were carried out using the GSAS suite of programs.

**Results and Discussion:**

We attempted to fit the 0.5 GPa data with the model of the ambient pressure phase but it was evident from the fit that the material had partially transformed into a different phase coexisting with the original tetragonal form. At 1.0 GPa, the splitting of the tetragonal {200} peaks make clear that there is an orthorhombic distortion of the original tetragonal cell with cell parameters 11.803 Å, 11.404 Å and 11.120 Å. This immediately suggests that the high pressure phase is isostructural with $Cs_4C_{60}$, a hypothesis that we have confirmed by detailed Rietveld refinement of the structure. We determined the space group to be *Immm* from considerations of systematic absences of Bragg reflections and the molecular symmetry of $C_{60}$. The loss of the four-fold rotation axis along [001] in this space group can be achieved either by aligning the 6:6 bonds in the *ab* plane with either the orthorhombic *a* or *b* axis. Comparing the quality of the refinements in either orientation we concluded that the 6:6 bonds perpendicular to c axis are aligned along the orthorhombic *b* axis (figure 2). The alternate orientation of the anion leads to unfavorable Rb-C contacts as short as 2.93 Å. Refinements were carried out keeping the $C_{60}$ molecule held rigid so that the radius of the ball is 3.54 Å and the single and double bonds are 1.45 Å and 1.39 Å respectively. The *Immm* space group has two in-equivalent sites for the cation. We were able to refine these Rb positional parameters without constraint confirming the choice of space group. The final two-phase Rietveld refinement at 10 kbar with $Rb_3C_{60}$ and $Rb_4C_{60}$ is shown in figure 3. A three-phase refinement, including an undistorted tetragonal phase at this pressure, did not improve the quality of the fit. All diffraction patterns measured for $Rb_4C_{60}$ can be fit with the same two coexisting phases, and a constant (~7 weight percent) fraction of $Rb_3C_{60}$, allowing us to extract the changes to the lattice parameters over this pressure range, as shown in Fig. 4. The contraction of the cell occurs along the crystallographic *b* axis that reduces the distance between the cation and the anion forcing the anion to lock into an ordered phase to avoid close contact with the Rb ions. Note that the *b* and *c* lattice parameters actually cross, although the symmetry remains orthorhombic. With increasing pressure the nearest neighbor distances in the orthorhombic phase decreases to substantially less than that in the cubic $Rb_3C_{60}$ phase.

The shortest Rb-C contacts are much smaller compared to $Cs_4C_{60}$: 3.13 Å(3.32 Å) and 3.15 Å(3.368 Å) to hexagon and pentagon atoms respectively. Such close contacts have also been observed in non-cubic superconducting fullerides such as $Ba_4C_{60}$ and $Sr_4C_{60}$. This evolution of inter-fullerene distances implies that the hopping of electrons and therefore the bandwidth should increase dramatically with increasing pressure.

As mentioned earlier recent x-ray diffraction patterns of $Rb_4C_{60}$ up to 5.20 GPa did not observe the orthorhombic structure described here[11]. We ascribe this discrepancy to the fact that the authors of that work tracked the positions of a few low order peaks instead of collecting and refining the entire diffraction pattern. It may be that the kink in lattice parameter *vs.* pressure observed in that work represents the tetragonal – orthorhombic transformation.

The insulating behavior of $A_4C_{60}$ system most likely originates from strong crystal field splitting of the $t_{1u}$ level arising from the less compact *bct* crystal structure or from J-

T distortion of the $C_{60}$ molecule. The interplay of the on-site coulomb repulsion and the band gap also play a significant role on the electronic states. NMR measurements find that the spin gap 0.15eV to the lowest triplet state in fact closes with the application of 10kbar pressure giving a finite DOS at the fermi level. Since the tetra-valent alkali fullerides are at the verge of the metal-insulator boundary, because of the narrow band nature of the $t_{1u}$-originated conduction band, minor distortions in the structure can easily alter their transport properties. The application of pressure lifts the orientational disorder and brings the buckyball much closer to each other than most doped fulleride systems. This most likely enhances electron hopping. The picture of the structural phase transition is consistent with the explanation that pressure increases the bandwidth in $Rb_4C_{60}$ thus driving the system to become conducting. The present work gives a clear picture of the structural behavior of $Rb_4C_{60}$ as a function of pressure. Measurements of optical properties of this system under pressure are clearly needed to have a better understanding of its transport properties. STM measurements have observed electron energy level shifts characteristic of a JT distortion of a monolayer film of $C_{60}^{4-}$ adsorbed on a surface, but the relative influence of single molecule *vs.* crystal field cannot be directly transferred from those experiments to bulk samples[12].

## Conclusion:

At this point, the transition pressure to the orthorhombic phase of $Rb_4C_{60}$ is not known; our diamond anvil cell measurements only set a lower limit of 0.4 GPa. Nor is the pressure of the transition to a conducting state known; there is only one data point at 0.8 GPa. Since the initial orthorhombic distortion is so small and orthorhombic symmetry appears to be associated with an insulating state of $Cs_4C_{60}$, we speculate that

$Rb_4C_{60}$ is also an insulator at the pressure where it first becomes orthorhombic, and that it becomes a conductor as the inter-fullerene distance is further reduced. It will be very interesting to measure transport properties, for example through far infrared reflectivity, as a function of pressure.

## Acknowledgement

We are grateful to Doug Robinson for crucial assistance at the MuCAT beamline of the Advanced Photon Source, and to Götz Bendele for experimental contributions in preliminary stages of this work. This work was partially supported by NSF grant DMR92-02528 and Department of Energy grant 6476789. Work at the National Synchrotron Light Source was partially supported by the Department of Energy under grant no. DE-FG02-86ER 45231. Use of the National Synchrotron Light Source, Brookhaven National Laboratory, was supported by the U.S. Department of Energy, Office of Science, Office of Basic Energy Sciences, under Contract No. DE-AC02-98CH10886.

**Figure Captions**

Figure 1: Evolution of synchrotron X-ray diffraction pattern of $Rb_4C_{60}$ as a function of applied pressure collected with sample inside a Diamond Anvil Cell. The data was collected using X-rays of wavelength 0.825Å. A small quantity of $Rb_3C_{60}$ is present in the starting sample.

Figure 2: The orientationally ordered *Immm* structure of $Rb_4C_{60}$ under pressure. The Rb1 cation is empty circles, Rb2 is shown as grey circle and carbon atoms are black. View along the (001) axis, showing the orientation of the 6:6 bond along the orthorhombic b-axis.

Figure 3: Rietveld refinement of $Rb_4C_{60}$ at 10kbar. The observed pattern is denoted by dots, the calculated and difference plots are shown in solid lines. Both peak markers for $Rb_3C_{60}$ and $Rb_4C_{60}$ are shown.

Figure 4: Upper plot: Nearest neighbor(C-C) distance of $Rb_4C_{60}$ as a function of pressure. For comparison these distances for the superconducting $Rb_3C_{60}$ is included. Lower plot: Dependence of the lattice parameters of $Rb_4C_{60}$ on pressure.

**Figure 1**

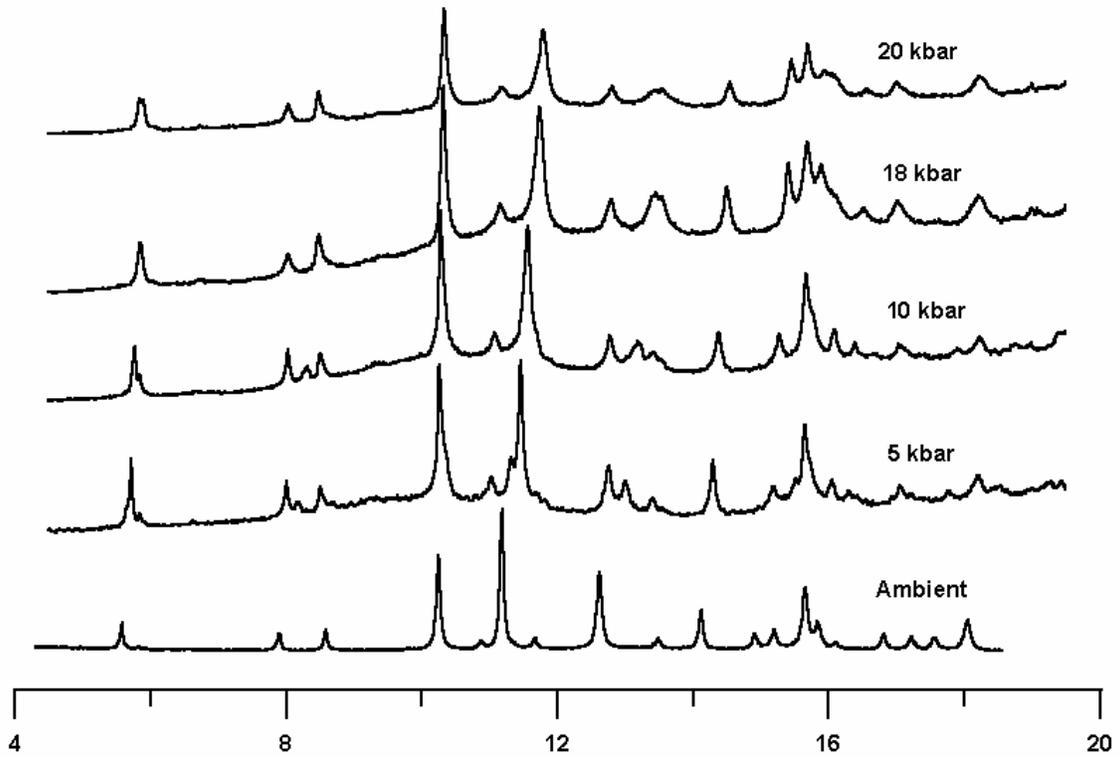

**Figure 2**

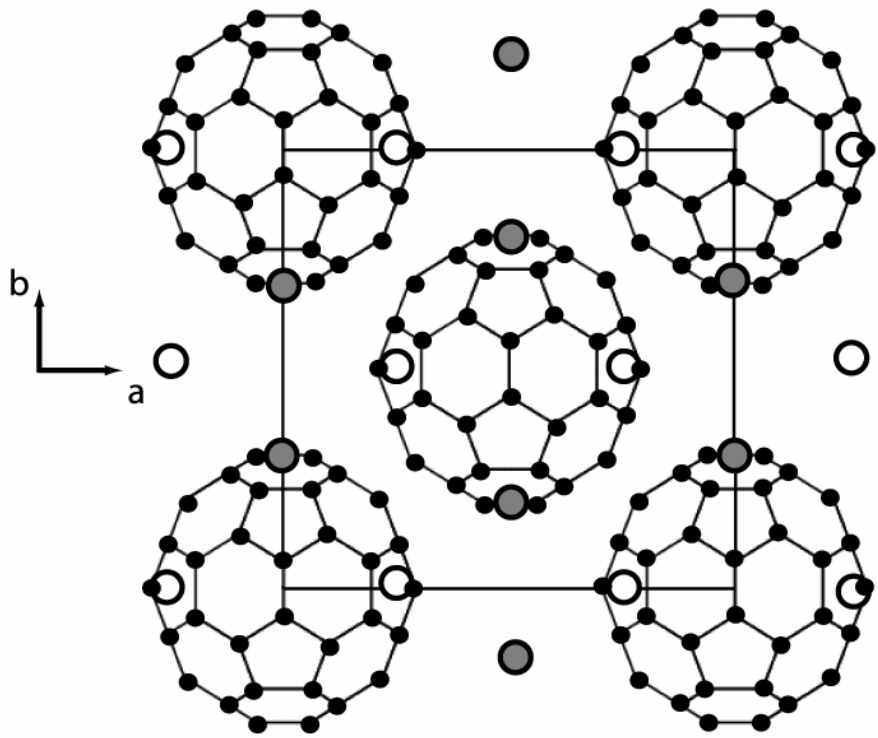

# Figure 3

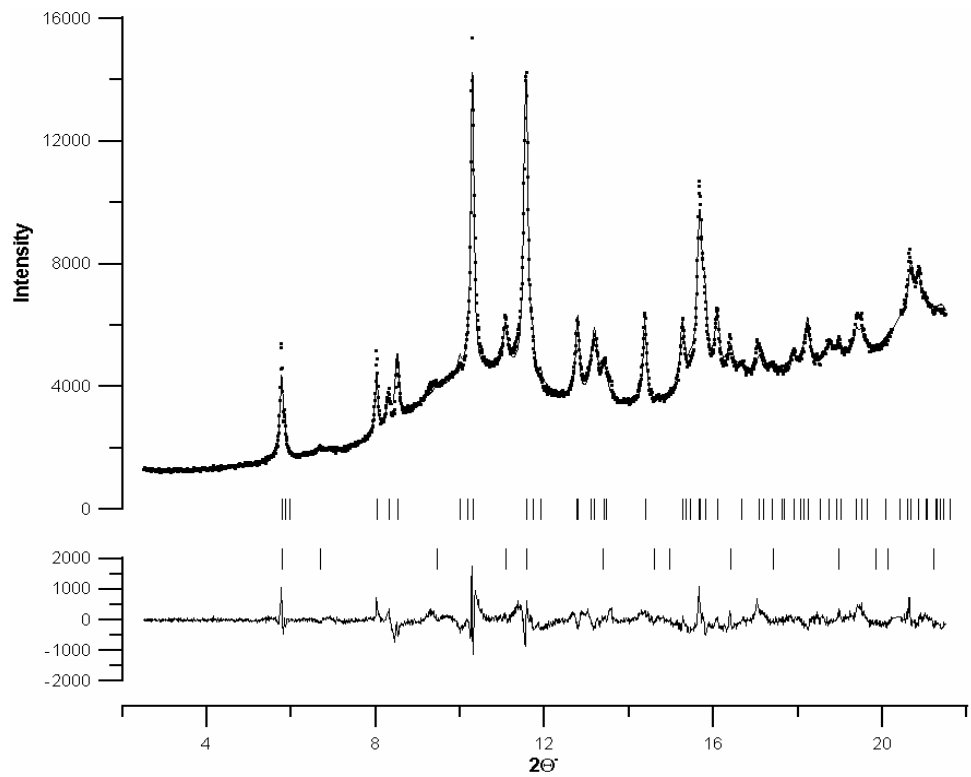

# Figure 4

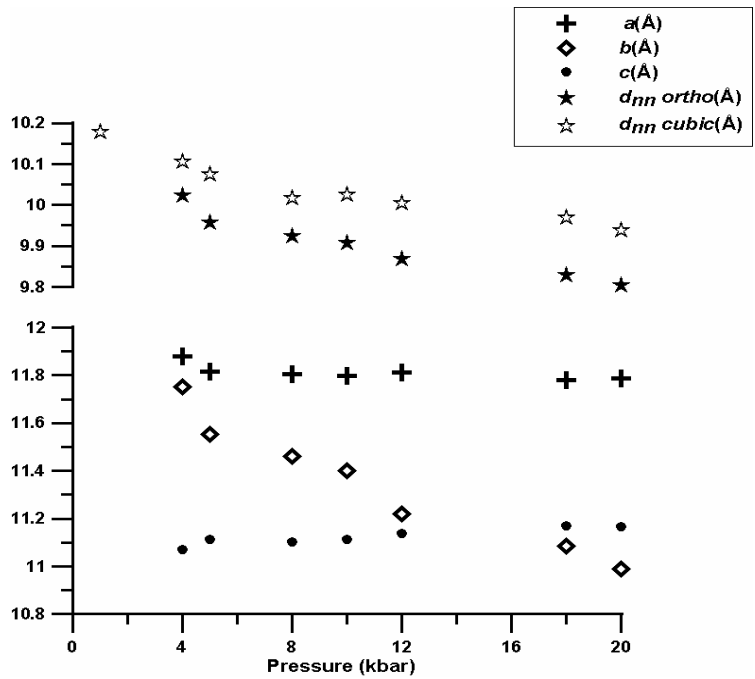